\documentclass[12pt]{article}

\long\def\symbolfootnote[#1]#2{\begingroup%
\def\thefootnote{\fnsymbol{footnote}}\footnote[#1]{#2}\endgroup}

 \font\tenrm=cmr10 \font\tenit=cmti10
\font\elevenbf=cmbx10 scaled\magstep 1 \font\elevenrm=cmr10
scaled\magstep 1  1

 \font\tenrm=cmr10 \font\tenit=cmti10
\font\elevenbf=cmbx10 scaled\magstep 1 \font\elevenrm=cmr10
scaled\magstep 1  1

\newcommand{\x}{\hat{x}}
\newcommand{\be}{\begin{equation}}
\newcommand{\bea}{\begin{eqnarray}}
\newcommand{\eea}{\end{eqnarray}}
\newcommand{\non}{\nonumber}
\newcommand{\ee}{\end{equation}}
\newcommand{\la}{\label}

\newcommand{\oal}{\overline{\alpha}}
\newcommand{\opi}{\overline{\pi}}
\newcommand{\ophi}{\overline{\phi}}
\newcommand{\D}{\mbox {D}}
\newcommand{\omu}{\overline{\mu}}
\newcommand{\oka}{\overline{\kappa}}
\newcommand{\osi}{\overline{\sigma}}
\newcommand{\otau}{\overline{\tau}}
\newcommand{\onu}{\overline{\nu}}
\newcommand{\obe}{\overline{\beta}}

\newcommand{\orho}{\overline{\rho}}
\newcommand{\oga}{\overline{\gamma}}
\newcommand{\oeps}{\overline{\epsilon}}
\newcommand{\ode}{\overline{\delta}}
\newcommand{\oio}{\overline{\iota}}

\newcommand{\oo}{\overline{o}}
\newcommand{\dA}{\dot{A}}
\newcommand{\dB}{\dot{B}}
\newcommand{\dC}{\dot{C}}

\newcommand{\dL}{\dot{L}}
\newcommand{\dE}{\dot{E}}
\newcommand{\dK}{\dot{K}}

\newcommand{\ola}{\overline{\lambda}}
\newcommand{\al}{\alpha}

\def\s={\stackrel{*}{=}}
\def\g0{\stackrel{\circ}{g}}
\def\half{\frac{1}{2}}

\def\x{x_1}
\def\y{x_2}
\def\xc{\overline{x_1}}
\def\yc{\overline{x_2}}

\def\square{\vcenter{\vbox{\hrule height.5pt \hbox{\vrule
width.5ptheight7pt \kern7pt \vrule width.5pt} \hrule height.5pt}}}
\renewenvironment{thebibliography}[1]
 { \elevenrm
   \begin{list}{\arabic{enumi}.}
    {\usecounter{enumi} \setlength{\parsep}{0pt}
     \setlength{\itemsep}{3pt} \settowidth{\labelwidth}{#1.}
     \sloppy
    }}{\end{list}}
\begin{document}

\newtheorem{theorem}{Theorem}
\newtheorem{lemma}{Lemma}

\begin{center}
\vglue 0.6cm

 {\elevenbf        \vglue 10pt
    Huygens' Principle for the Non-Self-Adjoint Scalar Wave \\
               \vglue 3pt
Equation on Petrov type III
Space-Times\symbolfootnote[2]{Published
on Ann. Inst. Henri Poincar\'e (A) Phys. Théorique \textbf{70}, 259 (1999).}}\\

\vglue 0.3cm { W. G. Anderson\\} \baselineskip=13pt {\tenit
Department of Physics, University of Alberta \\}
\baselineskip=12pt
{\tenit Edmonton, Alberta T6G 2J1, Canada}\\
\vglue 0.5cm {R. G. McLenaghan\\} \baselineskip=13pt {\tenit
Department of Applied Mathematics, University of Waterloo
\\} \baselineskip=12pt
{\tenit Waterloo, Ontario N2L 3G1, Canada}\\
\vglue 0.5cm
{\tenrm and }\\
\vglue 0.5cm { F. D.
Sasse\symbolfootnote[1]{fsasse@joinville.udesc.br}\\}
\baselineskip=12pt {\tenit Department of Mathematics, Centre for
Technological Sciences-UDESC
\\} \baselineskip=12pt {\tenit Joinville 89223-100, Santa
Catarina, Brazil } \vglue 0.8cm
\end{center}
{ \tenrm\baselineskip=11pt
 \noindent
\begin{quote}
{ Abstract} --- We prove that if the non-self-adjoint scalar wave
equation satisfies Huygens' principle on Petrov type III
space-times, then
 it is equivalent to the conformally invariant
 scalar wave equation.
\end{quote}
%\vspace{1cm}
\begin{quote}
\noindent { R\'esum\'e} --- On d\'emontre que la valit\'e du
principe de Huygens pour l'equation des onde scalaires
non-auto-adjointe sur un espace-temps g\'en\'eral de type III de
Petrov implique que l'equation est \'equivalente \'a l'equation
invariante conforme des ondes scalaires.
\end{quote}
}

\section{Introduction}
In this paper we consider the general
linear second-order hyperbolic equation
 on a four-dimensional curved space-time $V_4$ , with
$C^{\infty}$ coefficients:
\be
\label{1.1}
Pu:=g^{ab}\nabla_a\nabla_bu +A^a(x)\partial_au +C(x)u=0\,,
\ee
where $g^{ab}$ is a contravariant pseudo-Riemannian
metric with signature \linebreak
 $(+---)$ on $V_4$, u is the unknown scalar function,
$\nabla_a$ denotes the covariant derivative
with respect to the Levi-Civita connection, $A^a$ is a vector field
and $C$ is a scalar field in $V_4$.
All considerations are restricted
to a geodesically convex domain.
The equation (\ref{1.1}) is also
called  the {\it non-self-adjoint scalar
wave equation}.

{\it Huygens' principle}  is said to be valid for
an equation of the form
 (\ref{1.1}) if and only if for every Cauchy initial problem, and for each
point $x_0 \in V_4$, the solution depends only on the Cauchy data in an
arbitrarily small neighbourhood of $S \cap C^{-}(x_0)$, where $S$ denotes
the initial surface and $C^{-}(x_0)$ the past null conoid of $x_0$. Such
equations are called {\it Huygens' equations}.

Necessary conditions for the validity of
Huygens' principle for (\ref{1.1}) are
given by
\be
\la{conI}
(I)\qquad \qquad \qquad {\cal C}:=C-\half
A^i{}_{;i}-\frac{1}{4}A_iA^i-\frac{1}{6}R=0\,,
\ee
\be
\label{conII}
(II)\qquad \qquad  H^k{}_{a;k}=0\,,
\ee
\be
\label{conIII}
(III) \qquad S_{abk;}{}^{k}\,-\,
\frac{1}{2}C^{k}{}_{ab}{}^{l}L_{kl}=-5\left( H_{ak}H_{b}{}^k-\frac{1}{4}
g_{ab}H_{kl}H^{kl}\right)\,,
\ee
\be
\label{conIV}
(IV) \qquad TS(3S_{abk}H^k{}_c+C^k{}_{ab}{}^lH_{ck;l})=0\,,
\ee
\begin{eqnarray}
\nonumber
(V)\;&& TS\left(
3C^k{}_{ab}{}^l{}_{;}{}^mC_{kcdl;m}\,+\,8C^k{}_{ab}{}^l{}_{;c}
S_{kld}+40S_{ab}{}^k{}S_{cdk}-8C^k{}_{ab}{}^lS_{klc;d} \right.\\
\nonumber
&&-24C^k{}_{ab}{}^lS_{cdk;l}\,+\,
4C^k{}_{ab}{}^lC_l{}^m{}_{ck}L_{dm}+12C^k{}_{ab}{}^lC^m{}_{cdl}L_{km} \\
 &&
\non
\left.+12H_{ka;bc}H^k{}_d-16H_{ka;b}H^k{}_{c;d}-
84H^k{}_aC_{kbcl}H^l{}_d\right.\\
\label{conV}
&&\left.-18H_{ka}H^k{}_b L_{cd}\right)=0\,,
\end{eqnarray}
\begin{eqnarray}
\nonumber
(VI)\;&& TS\left(36C^k{}_{ab}{}^l
C_{lcdm;k}H^m{}_e - 6C^k{}_{ab}{}^l{}_{;c}
C_{lde}{}^mH_{km}-138S_{ab}{}^k C_{kcdl} H^l{}_e \right. \\
\nonumber
&&+6S_{abk}H^k{}_{c;de}+6C^k{}_{ab}{}^l{}_{;c}H_{kd;le}
-24S_{abk;c}H^k{}_{d;e}\\
\label{conVI}
&&\left. +12 C^k{}_{ab}{}^l L_{kc} H_{ld;e}
-9C^k{}_{ab}{}^l{}_{;c} L_{kd} H_{le}
-9S_{abk}L_{cd}H^k{}_e \right) =0 \,,
\end{eqnarray}
where
\begin{eqnarray}
\la{max}
&&H_{ij}:=A_{[i,j]}\,,\\
\la{weyl}
&&C_{ijkl}:=R_{ijkl}-2g_{[i[l}L_{j]k]}\,,\\
\la{S}
&&S_{ijk}:=L_{i[j;k]}\,,\\
\la{L}
&&L_{ij}:=-R_{ij}+\frac{R}{6} g_{ij}\,.
\end{eqnarray}
In the expressions above $R_{abcd}$ denotes the Riemann tensor,
$C_{abcd}$ the Weyl tensor, $R_{ab}:= g^{cd}R_{cabd}$, the Ricci
tensor, and $R:=g^{ab}R_{ab}$ the Ricci scalar associated to the
metric $g_{ab}$, and $A_a:=g_{ab}A^b$. Conditions $I-IV$ were
obtained by G\"unther \cite{gun52}, condition $V$ was derived by
W\"unsch \cite{wun70} for the self-adjoint case and McLenaghan
\cite{mcl74} for the non-self-adjoint case. Condition $VI$ was
obtained by Anderson and McLenaghan \cite{and94}. We shall use
here the two-component spinor formalism of Penrose \cite{pen60}
and the spin-coefficient formalism of Newman and Penrose
\cite{new62,pir64},  whose conventions we follow. The spinor
equivalents of the tensors (\ref{max}), (\ref{weyl}),
 (\ref{S}) (\ref{L}) are given by :
\bea
\la{Csp}
&&C_{abcd} \leftrightarrow
\Psi_{ABCD}\varepsilon_{\dot{A} \dot{B}} \varepsilon_{\dot{D} \dot{C}}
+\overline{\Psi}_{\dot{A}\dot{B}\dot{C}\dot{D}} \varepsilon_{AB} \varepsilon_{DC}\,,\\
\la{Hsp}
&&H_{ab} \leftrightarrow 2(\phi_{AB} \varepsilon_{\dA \dB} + \overline{\phi}_
{\dA \dB})\,,\\
\la{Lsp}
&&L_{ab} \leftrightarrow  2(\Phi_{A B \dot{A} \dot{B}} - \Lambda \varepsilon_{AB}
\varepsilon_{\dot{A} \dot{B}}) \,,\\
\la{Ssp}
&& S_{abc} \leftrightarrow  \Psi^D{}_{ABC;D \dA} \varepsilon_{\dC \dB}
+\overline{\Psi}^{D}{}_{\dA \dB \dC;D \dA} \varepsilon_{CB}\,,
\eea
where $\Psi_{ABCD}=\Psi_{(ABCD)}$ is the Weyl spinor,
$\Lambda=(1/24)R$,  $\phi_{AB}=\phi_{(AB)}$ is  the Maxwell spinor, and
$\Phi_{A B \dot{A} \dot{B}}=\Phi_{(A B) (\dot{A} \dot{B})}=\overline
{\Phi}_{A B \dot{A} \dot{B}}$ is  the trace-free Ricci
spinor.

We can set up, at each point of space-time, a {\it dyad basis}
$\{o^A, \iota^A\}$ satisfying the  relation
$o_A \iota^A=1$. The NP-components of
any spinor are defined by projecting
the spinor into the the local basis $\{o^A, \iota^A\}$. For the curvature
spinors and Maxwell spinor we have
\bea
\non
 \Psi_{ABCD}&=&\Psi_0\iota_{ABCD}-4\Psi_1 o_{(A}\iota_{BCD)}\
+6\Psi_2 o_{(AB}\iota_{CD)}\\
\label{3.26b}
&&-4\Psi_3o_{(ABC}\iota_{D)}
+\Psi_4o_{ABCD}\,,
\eea
\bea
\nonumber
\Phi_{AB\dot{A}\dot{B}}&=&\Phi_{22}o_{AB}\overline{o}_{\dot{A}\dot{B}}-
2\Phi_{21}o_{AB}\overline{o}_{\dot{(A}}\iota_{\dot{B)}}
-2\Phi_{12}o_{(A}\iota_{B)}\overline{o}_{\dot{A}\dot{B}}\\
\nonumber
&&+\Phi_{20}o_{AB}\overline{\iota}_{\dot{A}\dot{B}}+
\Phi_{02}\iota_{AB}\overline{\iota}_{\dot{A}\dot{B}}+
4\Phi_{11}o_{(A}\iota_{B)}\overline{o}_{\dot{(A}}\iota_{\dot{B)}}\\
\la{17}
&&-2\Phi_{10}o_{(A}\iota_{B)}\overline{o}_{\dot{(A}}\iota_{\dot{B)}}-
2\Phi_{01}\iota_{AB}\overline{\iota}_{\dot{A}\dot{B}}\,,
\end{eqnarray}
\be
\la{18}
\phi_{AB}=\phi_0\iota_{AB}-2\phi_1 o_{(A}\iota_{B)}+\phi_2 o_{AB}\,,
\ee
where $\iota_{ABCD}=\iota_A\iota_B\iota_C\iota_D$, etc.

The covariant derivatives of the  dyad basis spinors are given in terms
of the NP-spin coefficients by
\be
\label{3.29}
o_{A;B \dot{B}}= o_A I_{B \dot{B}} + \iota_A II_{B \dot{B}}\,,\;\;\;\;
\iota_{A;B \dot{B}}= o_A III_{B \dot{B}} - \iota_A I_{B \dot{B}}\,,
\ee
where
\be
\label{3.30}
\left.\begin{array}{l}
I_{B \dot{B}}:= \gamma o_B \overline{o}_{\dot{B}} -
\al o_B \overline{\iota}_{\dot{B}} - \beta \iota_B \overline{o}_{\dot{B}}
+ \varepsilon  \iota_B \overline{\iota}_{\dot{B}}\,,\\
II_{B \dot{B}}:= -\tau o_B \overline{o}_{\dot{B}} +
\rho o_B \overline{\iota}_{\dot{B}} + \sigma \iota_B \overline{o}_{\dot{B}}
-\kappa  \iota_B \overline{\iota}_{\dot{B}}\,,\\
III_{B \dot{B}}:= \nu o_B \overline{o}_{\dot{B}} -
\lambda o_B \overline{\iota}_{\dot{B}} - \mu \iota_B \overline{o}_{\dot{B}}
+ \pi  \iota_B \overline{\iota}_{\dot{B}}\,.
\end{array}\right\}
\ee

The necessary conditions $I$ - $VI$ can be expressed in terms of
dyad components, containing only the Newman-Penrose scalars. This
conversion procedure consists in two steps. Firstly, we have to
convert the tensorial expressions into  spinor form, using
(\ref{Csp})-(\ref{Ssp}). Secondly, the spinor equations must be
expressed in terms of  dyad components, using (\ref{3.26b}),
(\ref{17}) (\ref{18}), and (\ref{3.29}). We perform these lengthy
calculations automatically with the  {\tt NPspinor} package
\cite{cza87,cza92}, available in the Maple computer algebra
system.

Using the fact that $H_{[ij;k]}=0$, it can be shown that
the spinor form of condition $II$ is given by:
\be
\la{eq-IIs}
(IIs) \qquad \phi_{AK;}{}^K{}_{\dA}=0\,.
\ee
For condition $III$ a direct application of the correspondence relations
yields
\bea
\non
&&\Psi_{ABKL;}{}^K{}_{\dA}{}^L{}_{\dB}+ \overline{\Psi}_{\dA \dB \dK \dL;}{}
^{\dK}{}_A{}^{\dL}{}_B+\Psi_{AB}{}^{KL} \Phi_{K L \dA \dB}\\
\la{IIIs}
&&\qquad+\overline{\Psi}_{\dA \dB}{}^{\dK \dL}\Phi_{\dK \dL A B}
+10 \phi_{AB}\ophi_{\dA \dB}=0\,.
\eea
Instead of (\ref{IIIs}) we shall use a stronger form of this condition, obtained
by W\"unsch \cite{wun89} and McLenaghan and Williams \cite{mcl90}:
\be
\la{IIIs2}
(IIIs) \qquad \nabla^K{}_{\dA}\nabla^L{}_{\dB}
\Psi_{ABLK}+\Phi^{KL}{}_{\dA\dB}\Psi_{ABKL}
+5\phi_{AB}\overline{\phi}_{\dA\dB}=0\,.
\ee
While the original necessary condition (\ref{IIIs}) is Hermitian, (\ref{IIIs2})
is  complex.

The conversion of the remaining conditions to the respective
spinor form and  the determination of the dyad components is done
automatically by defining templates in the {\tt NPspinor} package.
The spinor form of the trace-free symmetric part of a tensor is
obtained by  taking the correspondent spinor equivalent of that
tensor and symmetrizing with respect to all dotted and undotted
indices \cite{wal88}.  McLenaghan and Walton \cite{mcl88} have
shown that any non-self-adjoint equation (\ref{1.1}) on any Petrov
type N space-time satisfies Huygens' principle if and only if it
is equivalent to a scalar equation with $A_a=0$ and $C=0$, on an
space-time corresponding to the exact plane-wave metric.

In this paper we prove the correspondent result for Petrov type III space-times.
In this case the Weyl spinor has the form:
\be
\Psi_{ABCD}=\al_{(A}\al_B \al_C \beta_{D)}\,,
\ee
where $\al_A$ and $\beta_A$ are the principal spinors. If we choose the spin basis
such that $\al_A$ is proportional to the dyad basis spinor $ o_A$ and
$\beta$ proportional to $\iota_A$, we obtain from (\ref{3.26b}):
\be
\label{4.83}
\Psi_{ABCD}=-4\Psi_3 o_{(ABC}\iota_{D)}\,.
\ee
The dyad transformation
\be
\label{3.37}
o'=e^{w/2}o\,,\;\;\; \iota' =e^{-w/2}(\iota+q o)\,,
\ee
where $w$  and $q$ are complex functions, induces the following transformation on
$\Psi_3$:
\be
\Psi'_3=e^{-w}(\Psi_3+3q\Psi_2+3q^2\Psi_1+q^3\Psi_0)\,.
\ee
Thus, we can choose the tetrad such that
$\Psi_3=-1$, so that
\be
\label{4.84}
\Psi_{ABCD}=4o_{(ABC}\iota_{D)}\,.
\ee

In a  previous paper,  Anderson, McLenaghan and Walton
\cite{and96} have proved the following theorems:
\begin{theorem}
\la{teo-4.1}
The validity of Huygens' principle for any
non-self-adjoint scalar wave equation (\ref{1.1}) in any Petrov type III space-time
implies that the space-time
is conformally related to one in which every repeated null vector field
of the Weyl tensor $l_a$, is recurrent, i.e.,
\be
\label{teo1.1}
l_{[a}l_{b;c]}=0\,.
\ee
\end{theorem}
\begin{theorem}
\la{teo-4.2}
There exist no  non-self-adjoint Huygens' equations (\ref{1.1})
 on any Petrov type III space-time for which
the following conditions hold
\be
\label{teo-5.2}
\left.\begin{array}{l}
\Psi_{ABCD;E\dE}\iota^A\iota^B\iota^C o^D \iota^E \oo^{\dE}=0\,,\\
\Psi_{ABCD;E\dE}\iota^A\iota^B\iota^C o^D o^E \oo^{\dE}=0\,,\\
\Psi_{ABCD;E\dE}\iota^A\iota^B\iota^C \iota^D o^E \oo^{\dE}=0\,.
\end{array}\right\}
\ee
\end{theorem}
In the next section we show that the restrictions imposed by
 these two theorems can be removed.

\section{Main Theorem}
The main result of this paper is expressed by the theorem:
\begin{theorem}[Main Theorem]
\la{teo-4.3}
If a  non-self-adjoint scalar wave equation of the form (\ref{1.1})
satisfies Huygens' principle on any
 Petrov type III space-time, then it must be
equivalent to a conformally invariant scalar wave equation
\be
\la{eq-conf}
g^{ab}\nabla_a\nabla_bu+\frac{1}{6}Ru=0\,.
\ee
\end{theorem}
%%%%%%%%%%%%%%%%%%%%%%%%%%%%%%%%%%%%%%%%%%%%%%%%%%%%%%%%%%%%%%%%%%%%%%%%%%%%%%%%%%%%
{\bf Proof}

We shall  first prove, using the necessary conditions
$II$ to $VI$ given by (\ref{conII}) -   (\ref{conVI}),
  that the assumption $A_{[a,b]}:=H_{ab}\neq 0$
leads to a contradiction. In terms of the dyad components of the Maxwell
tensor $H_{ab}$, this is the same as proving that the necessary conditions imply
$\phi_0=\phi_1=\phi_2=0$. Finally we invoke a lemma by G\"unther
\cite{gun52} that states that {\it every equation of the form (\ref{1.1}) for which
$A_{[a,b]}:=0$ is related by a trivial transformation to one for which $A_a=0$}.
It then follows from the necessary condition $I$ (eq. (\ref{conI})) that $C=R/6$.
We  use  a notation for the dyad components of the necessary
conditions in the form $X_{ab}$, where X is
the Roman numeral corresponding to
the necessary condition, $a$ denotes the number of indices corresponding to
the dyad spinor $\iota$ and $b$ the number of dotted indices corresponding
to the dyad spinor $\oio$. We shall refer to the Newman-Penrose field equations
using the notation NP1, NP2, etc. as listed in the Appendix.
The methods employed in this proof are similar to those
used in \cite{mcl96}.

We start with the result obtained by
 Anderson and McLenaghan \cite{and96},
expressed in following lemma :
\begin{lemma}
\la{lemma4.1}
 For the non-self-adjoint scalar equation of the form ({\ref{1.1}), in
Petrov type III space-times, the necessary conditions}
$II$, $III$, $IV$, $V$ and $VI$ together with the assumption
that the Maxwell spinor $\phi_{AB}$ is nonzero, imply
that there exists a spinor dyad $\{o_A,\iota_A\}$
and a conformal transformation such that
\be
\label{5.84}
\left.\begin{array}{l}
\kappa=\sigma=\rho=\tau=\epsilon=0\,,\\
\Psi_0=\Psi_1=\Psi_4=0\,,\;\Psi_3=-1\,,\\
\Phi_{00}=\Phi_{01}=\Phi_{02}=\Lambda=0\,,\\
\D \al=\D \beta = \D \Phi_{11}=0\,.
\end{array}\right\}
\ee
\end{lemma}
In what follows we shall use the relations (\ref{5.84}) where
necessary. It was shown in \cite{and96} that $\phi_0=\phi_1=0$.
Thus, what remains to be proved is that the assumption
$\phi_2\neq0$ leads to a contradiction. Let us assume initially
that $\al\beta\pi \neq 0$. From $II_{01}$, $II_{00}$, $III_{10}$,
(NP6) and (NP25) we have, respectively, \bea \label{5.85}
&&\D\phi_2=0\,,\\
\label{5.86}
&&\delta \phi_2=-2\phi_2\beta\,,\\
\label{5.87}
&&\delta \beta=-\beta(\oal+\beta)\,,\\
\label{5.88}
&&\D \gamma=\al\opi+\beta\pi+\Phi_{11}\,,\\
\label{5.89}
&&\delta \Phi_{21}=2(\al+2\pi+\lambda\Phi_{11}-\al\Phi_{21})\,.
\eea
By adding (NP22) to the complex conjugate of (NP23), and solving for
$\D\Phi_{12}$ we get
\be
\label{5.90}
\D \Phi_{12}=2\opi\Phi_{11}\,,
\ee
and
\be
\label{5.91}
\delta \Phi_{11}=0\,.
\ee
Subtracting (NP24) from (NP29) and solving for $\ode \Phi_{12}$ we obtain
\be
\label{5.92}
\ode \Phi_{12}=2(-\beta+\Phi_{11} \omu-\obe\Phi_{12})\,,
\ee
Adding (NP24) to two times (NP29) and solving for $\D\Phi_{22}$ we get
\be
\label{5.93}
\D\Phi_{22}=2(-\beta-\obe+\Phi_{21}\opi+\Phi_{12}\pi)\,.
\ee
By substituting (\ref{5.86}) into $IV_{10}$ we find
\be
\label{5.94}
\ode \phi_2=2(3\obe\phi_2-\al\phi_2+2\pi\ophi_2+\al\ophi_2)\,.
\ee
Using (\ref{5.93}) and (\ref{5.94}), $V_{20}$ and $VI_{03}$ can
be written respectively as
\bea
\nonumber
&&-24\ophi_2\phi_2\beta^2+\phi_2{}^2(-6\oal\opi+18\beta\opi
+9\beta\oal-\epsilon\oal^2-2\delta\opi-\delta \oal)\\
\label{5.95}
&&+12\oal^2+24\opi\oal+80\opi\beta++
4\delta\oal+8\delta\opi+44\beta\oal=0\,,
\eea
\be
\la{5.96}
\beta(\phi_2(\delta \oal+2\delta\opi+3\oal^2-18\opi\beta+6\oal\opi-9\oal\beta)+
24\ophi_2\beta^2)=0\,.
\ee
Eliminating  $\delta(\oal+2\pi)$ from (\ref{5.95}) and (\ref{5.96}) we find
\be
\la{5.98a}
\frac{\beta^2}{\phi_2{}^2-4}\left(19\opi\phi_2+10\oal\phi_2-12\beta\ophi_2\right)=0\,.
\ee
The denominator $-\phi_2{}^2+4$ in the expression above must be nonzero,
since $\phi_2$ cannot be constant. Otherwise, from (\ref{5.86}), we would
have $\beta=0$.

In order to determine further side relations we still need to find the Pfaffians
$\delta \al$, $\delta \obe$, and $\delta \pi$, in terms of
$\ode \al$. From (NP6), (NP7), (NP8), (NP9) and (NP12) we have
\bea
\la{5.97}
&&\ode\pi=\D \lambda-\pi^2-\pi\al+\pi\obe\,,\\
\la{5.98}
&&\delta \pi=\D\mu-\pi\opi+\pi\oal-\beta\pi\,,\\
\la{5.99}
&&\D\nu=\Delta\pi+\pi\mu+\pi\mu+\opi\lambda+\pi\gamma-\pi\oga-1+\Phi_{21}\,,\\
\la{5.100}
&&\delta \al=\ode \beta+\al\oal+\beta\obe-2\obe\oal+\Phi_{11}\,.
\eea
Using (\ref{5.90}) and  (\ref{5.93}),  we now evaluate the NP commutator
$[\delta,\D]\Phi_{22}-[\Delta,\D]\Phi_{12}$ to obtain
\begin{eqnarray}
\nonumber
&&  - 2\delta\obe + \D\Delta
\Phi_{12} + 2\opi\Phi_{11}\oga + 2\beta\obe
 + 4\beta^{2} + 2\pi\ola\Phi_{11} - 6\beta\Phi_{21}
\opi + 4\pi\opi \\
\nonumber
 & & \mbox{} + 4\beta\oal - \D\delta\Phi_{22}
 - 4\oal\Phi_{12}\pi - 2\beta\Phi_{12}\pi - 2
\oal\Phi_{21}\opi + 2\opi\Phi_{11}\gamma - 2
\opi\Phi_{11}\omu \\
\nonumber
 & & \mbox{} + 4\opi\Phi_{11}\mu + 2\opi\Phi_{12}
\pi + 2\Phi_{21}\delta\opi + 2\Phi_{12}
\delta\pi - 2\Phi_{11}\Delta\opi + 2
\oal\obe - 6\opi\obe \\
\la{5.101}
 & & \mbox{} + 2\Phi_{21}\opi^{2} + 2\opi\Phi_{12}\obe+ 2\pi\oal=0\,.
\end{eqnarray}

 From $[\Delta,\D]\Phi_{12}$ we get
\begin{eqnarray}
\nonumber
&&\D\Delta\Phi_{12} = 2\Phi_{11}
\Delta\opi - 2\opi\Phi_{11}\gamma - 2\opi
\Phi_{11}\oga + 2\pi\oal + 4\pi\opi + 2
\pi\ola\Phi_{11}  \\
\la{5.102}
&& \qquad- 2\oal\Phi_{12}\pi  - 2\opi\beta + 2\opi\Phi_{11}
\omu - 2\opi\Phi_{12}\obe\,.
\end{eqnarray}
From (NP26),
\be
\la{5.103}
\delta \Phi_{22}=\Delta\Phi_{12}+2\oga+4\omu-2\onu\Phi_{11}+2\ola\Phi_{21}
+2\Phi_{12}\oga+2\Phi_{12}\mu-2\Phi_{22}\beta-2\Phi_{22}\oal\,.
\ee
Substituting (\ref{5.103}), (\ref{5.102}), (\ref{5.88}),
 (\ref{5.97}), (\ref{5.98}) and (\ref{5.99}) into (\ref{5.101})
we have
\be
\la{5.104}
\delta \obe=-\oal\obe-4\opi\obe-2\D\omu-\beta\obe+2\pi\opi-2\Phi_{11}\,.
\ee
Now, using (\ref{5.85}), (\ref{5.86}), (\ref{5.94}),  (\ref{5.98}), (\ref{5.100}) and
(\ref{5.104})
in the commutator $[\ode,\delta]\phi_2$,
and solving for $\D \mu $ we obtain
\begin{eqnarray}
\nonumber
&&\D\mu = -  \frac {1}{12\ophi_2}(8
\ophi_2\al\opi + 24\ophi_2\beta\pi - 8\phi_2
\al\opi + 10\ophi_2\Phi_{11} + 12\beta\ophi_2
\al - 12\oal\phi_2\obe \\
\la{5.105}
 & &  + 4\ophi_2\pi\opi + 4\oal\ophi_2
\al - 4\oal\phi_2\al - 2\phi_2\Phi_{11}
 - 24\phi_2\opi\obe + 8\ophi_2\pi\oal)\,.
\end{eqnarray}
The  explicit form of $\D\lambda$ can be determined from (\ref{5.95}),
(\ref{5.96}) and (\ref{5.97}), and is given by:
\bea
\nonumber
&&\D\lambda=\frac{1}{2(-\ophi_2{}^2+4)}\left(\ophi_2{}^2(3\al^2-2\pi^2
+4\al\pi-16\obe\pi-9\obe\al+\ode\al)\right.\\
\la{5.106}
&&\left.+8\pi^2 -16\pi\al-12\al^2+24\ophi_2\phi_2\obe^2-88\pi\obe-44\obe\al
-4\ode\al)\right)\,.
\eea
Finally, from $V_{12}$
 we get
\be
\la{5.107a}
\D\pi=0\,.
\ee

We have now determined all the Pfaffians needed for finding new side relations using
integrability conditions. Before  proceeding we go back to (\ref{5.98a}) and introduce
a further simplification by expressing $\phi_2$ in terms of $\ophi_2$,
\be
\la{5.107}
\phi_2=\frac{\ophi_2}{12\obe}(19\pi+10\al)\,.
\ee
Substituting (\ref{5.107})
into the numerator of the complex conjugate of (\ref{5.98a}),
 we find
\be
\la{5.108}
S_1:=361\pi\opi+190\al\opi+190\pi\oal+100\al\oal-144\beta\obe=0\,.
\ee
Another side relation can be determined from the NP commutator
$[\ode\,,\,\delta](\al+2\pi)$:
\begin{eqnarray}
\nonumber
&& ( - 820\,\al^{2}\,\Phi_{11}
 + 95\,\pi^{2}\,\Phi_{11} + 2680\,\al^{3}\,\oal - 5448\,\al\,
\obe\,\Phi_{11} - 10236\,\pi\,\obe\,\Phi_{11} \\
\nonumber
 & & \mbox{} + 5360\,\al^{3}\,\opi - 9648\,\obe\,\al^{2}
\,\beta - 38016\,\pi^{2}\,\beta\,\obe - 13926\,\pi^{2}\,\obe
\,\oal - 2508\,\pi^{2}\,\opi\,\obe \\
\nonumber
 & & \mbox{} - 3816\,\obe\,\al^{2}\,\oal - 720\,\al^{2}\,
\opi\,\obe + 61628\,\al\,\pi^{2}\,\opi + 40128\,\pi^{
3}\,\opi + 20064\,\pi^{3}\,\oal \\
\nonumber
 & & \mbox{} + 30814\,\pi^{2}\,\al\,\oal + 15752\,\pi\,\al^{2
}\,\oal - 38376\,\pi\,\obe\,\al\,\beta + 31504\,\pi\,\al^{2
}\,\opi \\
\la{5.125}
 & & \mbox{} - 14592\,\al\,\pi\,\obe\,\oal - 2688\,\pi\,
\al\,\opi\,\obe - 1508\,\pi\,\al\,\Phi_{11}) \left/
{\vrule height0.37em width0em depth0.37em} \right. \! \!
(\,19\,\pi + 10\,\al\,)=0.
\end{eqnarray}
It follows from (\ref{5.107}) that the numerator $19\,\pi + 10\,\al$ in the
preceding equation must be non-zero.
Solving this equation for $\Phi_{11}$ we obtain
\begin{eqnarray}
\nonumber
\lefteqn{\Phi_{11} = (2680\,\al^{3}\,\oal + 5360\,\al^{3}
\,\opi - 9648\,\obe\,\al^{2}\,\beta - 38016\,\pi^{2}\,\beta\,
\obe - 13926\,\pi^{2}\,\obe\,\oal} \\
\nonumber
 & & \mbox{} - 2508\,\pi^{2}\,\opi\,\obe - 3816\,\obe
\,\al^{2}\,\oal - 720\,\al^{2}\,\opi\,\obe + 61628\,\al\,
\pi^{2}\,\opi + 40128\,\pi^{3}\,\opi \\
\nonumber
 & & \mbox{} + 20064\,\pi^{3}\,\oal + 30814\,\pi^{2}\,\al\,
\oal + 15752\,\pi\,\al^{2}\,\oal - 38376\,\pi\,\obe\,
\al\,\beta \\
\nonumber
 & & \mbox{}+ 31504\,\pi\,\al^{2}\,\opi - 14592\,\al\,\pi\,\obe\,\oal
- 2688\,\pi\,\al\,
\opi\,\obe) \left/ ( 820\,\al^{2} - 95\,\pi^{2}  \right. \\
\la{5.127}
 & &   + 5448\,\obe\,\al + 10236
\,\pi\,\obe + 1508\,\pi\,\al\,)\,,
\end{eqnarray}
where, for now, we assume that the denominator in the expression
above, \be \la{5.128}
d_1:=820\al^2-95\pi^2+5448\obe\al+10236\pi\obe-1508\pi\al\,, \ee
is non-zero. Evaluation of $\delta\phi_2+2\phi_2\beta=0$ (cf. Eq.
(\ref{5.90})), using (\ref{5.107}), and solving for $\Phi_{11}$
gives
\begin{eqnarray}
\nonumber
\lefteqn{\Phi_{11} = (700\,\al^{2}\,\oal + 5300\,\al\,\pi
\,\opi + 2650\,\pi\,\al\,\oal + 2508\,\pi^{2}\,\oal
 + 1400\,\al^{2}\,\opi} \\
\nonumber
 & & \mbox{} - 2520\,\obe\,\al\,\beta - 4752\,\pi\,\beta\,\obe
- 1650\,\pi\,\obe\,\oal - 132\,\pi\,\opi\,\obe
 + 5016\,\pi^{2}\,\opi \\
\la{5.129}
 & & \mbox{} - 900\,\obe\,\al\,\oal - 72\,\al\,\opi\,
\obe) \left/ {\vrule height0.37em width0em depth0.37em}
 \right. \! \! (\,-437\,\pi + 372\,\obe - 230\,\al\,)\,,
\end{eqnarray}
where we assume, for now, that the denominator of (\ref{5.129}), given by
\be
\la{5.130}
d_2:=437\pi-372\obe+230\al \neq 0\,,
\ee
is nonzero.
Evaluating the commutators $[\ode\,,\,\delta]\obe$ and $[\ode\,,\,\delta]\ophi_2$, and
 solving each one for
$\ode \al$ we get, respectively,
\begin{eqnarray}
\nonumber
\lefteqn{\ode\al =( - 308\,\pi\,\al\,\Phi_{11} + 8016\,\pi
\,\al^{2}\,\opi + 234\,\obe\,\al^{2}\,\beta + 3972\,\pi\,
\al^{2}\,\oal + 78\,\obe\,\al^{2}\,\oal} \\
\nonumber
 & & \mbox{} + 759\,\al\,\obe\,\Phi_{11} - 29\,\al^{2}\,\Phi_{11}
+ 1440\,\al^{3}\,\opi + 702\,\al^{3}\,\oal + 1386\,\pi\,\obe\,\Phi_{11} \\
\nonumber
 & & \mbox{} - 513\,\pi^{2}\,\Phi_{11} + 1296\,\pi\,\obe\,\al
\,\beta + 528\,\pi^{2}\,\obe\,\oal + 14872\,\al\,\pi^{2}\,
\opi + 7436\,\pi^{2}\,\al\,\oal \\
\nonumber
 & & \mbox{} + 1584\,\pi^{2}\,\beta\,\obe + 4598\,\pi^{3}\,
\oal + 9196\,\pi^{3}\,\opi + 432\,\al\,\pi\,\obe\,
\oal - 108\,\pi\,\al^{2}\,\beta  \\
\la{5.131}
 & & - 54\,\al^{3}\,\beta)  / (\,  12\,\pi\,\oal + 6\,\al\,\oal +
 36\,\beta\,\pi + 18\,\beta\,\al + 3\,\Phi_{11}\,)\,,
\end{eqnarray}
\be
\la{5.132}
\ode\al =  - \,{\displaystyle \frac {6\,\al^{2}\,\opi -
572\,\pi\,\opi\,\obe - 286\,\pi\,\obe\,\oal - 157
\,\obe\,\al\,\oal + 3\,\al^{2}\,\oal - 314\,\al\,
\opi\,\obe}{\oal + 2\,\opi}}\,,
\ee
where we assume, for the moment, that the denominators of (\ref{5.131}) and
(\ref{5.132}), given by
\bea
\la{5.133}
&&d_3:=4\pi\oal+2\al\oal+12\beta\pi+6\beta\al+\Phi_{11} \,,\\
\la{5.134}
&&d_4:=\al+2\pi\,,
\eea
are nonzero.
By subtracting (\ref{5.131}) from (\ref{5.132}) and solving for $\Phi_{11}$
we have
\begin{eqnarray}
\nonumber
\lefteqn{\Phi_{11} = ( - 864\,\obe\,\al^{2}\,\oal +
7436\,\pi^{2}\,\al\,\oal - 3168\,\al\,\pi\,\obe\,\oal
 + 4008\,\pi\,\al^{2}\,\oal} \\
\nonumber
 & & \mbox{} - 2904\,\pi^{2}\,\obe\,\oal + 720\,\al^{3}\,
\oal + 9196\,\pi^{3}\,\opi - 8712\,\pi^{2}\,\beta\,\obe
 + 14872\,\al\,\pi^{2}\,\opi \\
\nonumber
 & & \mbox{}+ 4598\,\pi^{3}\,\oal - 2592\,\obe\,\al^{2}\,\beta + 8016\,\pi\,\al^{2}
\,\opi + 1440\,\al^{3}\,\opi - 9504\,\pi\,\obe\,\al\,
\beta)  \\
\la{5.135}
 & & \left/(-288\,\obe\,\al + 20\,\al^{2} - 528\,\pi\,\obe + 513
\,\pi^{2} + 308\,\pi\,\al\,)\right.\,,
\end{eqnarray}
where the denominator of (\ref{5.135}), \be \la{den4.5}
d_5:=-288\,\obe\,\al + 20\,\al^{2} - 528\,\pi\,\obe + 513
\,\pi^{2} + 308\,\pi\,\al\,, \ee is assumed to be nonzero for the
moment.
Subtracting (\ref{5.127}) from (\ref{5.129}) and taking
the numerator we find
\begin{eqnarray}
\nonumber
\lefteqn{S_2 := 39914208\,\pi^{2}\,\al^{2}\,\opi +
19957104\,\pi^{2}\,\al^{2}\,\oal + 35332704\,\pi^{3}\,\opi\,\obe} \\
\nonumber
 & & \mbox{} + 12279168\,\pi^{3}\,\obe\,\oal + 1190400\,
\oal\,\al^{4} - 12773376\,\obe^{2}\,\oal\,\pi\,\al -
3483648\,\obe^{2}\,\oal\,\al^{2} \\
\nonumber
 & & \mbox{} + 1200960\,\obe\,\oal\,\al^{3} - 11708928\,
\obe^{2}\,\oal\,\pi^{2} - 20739456\,\pi\,\obe\,\al^{2
}\,\beta \\
\nonumber
 & & \mbox{} + 8008704\,\al^{2}\,\pi\,\obe\,\oal +
30335616\,\pi\,\al^{2}\,\opi\,\obe + 56708640\,\pi^{2}\,
\al\,\opi\,\obe \\
\nonumber
 & & \mbox{} - 32440608\,\pi^{2}\,\obe\,\al\,\beta - 16161552\,
\pi^{3}\,\beta\,\obe + 8022720\,\pi\,\al^{3}\,\oal +
5408640\,\al^{3}\,\opi\,\obe \\
\nonumber
 & & \mbox{} + 16045440\,\pi\,\al^{3}\,\opi + 8529708\,\pi^{4
}\,\oal + 43221504\,\al\,\pi^{3}\,\opi + 21610752\,\pi^{3
}\,\al\,\oal \\
\nonumber
 & & \mbox{} + 17059416\,\pi^{4}\,\opi + 2380800\,\al^{4}\,
\opi + 17343792\,\al\,\pi^{2}\,\obe\,\oal - 10139904
\,\obe^{2}\,\al^{2}\,\beta \\
\nonumber
 & & \mbox{} - 124416\,\al^{2}\,\opi\,\obe^{2} - 418176\,
\pi^{2}\,\opi\,\obe^{2} - 456192\,\pi\,\al\,\opi\,
\obe^{2} - 4285440\,\obe\,\al^{3}\,\beta \\
\la{5.136}
 & & \mbox{} - 37407744\,\pi\,\obe^{2}\,\al\,\beta - 34499520\,
\pi^{2}\,\beta\,\obe^{2}=0\,.
\end{eqnarray}
By subtracting (\ref{5.129}) from (\ref{5.135}) and taking the numerator,  we
obtain a third side relation:
\begin{eqnarray}
\nonumber
\lefteqn{S_3 :=  - 9374472\,\pi^{2}\,\al^{2}\,\opi -
4687236\,\pi^{2}\,\al^{2}\,\oal + 6137076\,\pi^{3}\,\opi
\,\obe + 5150178\,\pi^{3}\,\obe\,\oal} \\
\nonumber
 & & \mbox{} - 179600\,\oal\,\al^{4} - 2128896\,\obe^{2}
\,\oal\,\pi\,\al - 580608\,\obe^{2}\,\oal\,\al^{2} +
686160\,\obe\,\oal\,\al^{3} \\
\nonumber
 & & \mbox{} - 1951488\,\obe^{2}\,\oal\,\pi^{2} + 4189824
\,\pi\,\obe\,\al^{2}\,\beta + 4040184\,\al^{2}\,\pi\,\obe\,
\oal \\
\nonumber
 & & \mbox{} + 5272368\,\pi\,\al^{2}\,\opi\,\obe +
9852984\,\pi^{2}\,\al\,\opi\,\obe + 8913384\,\pi^{2}\,
\obe\,\al\,\beta \\
\nonumber
 & & \mbox{} + 6244920\,\pi^{3}\,\beta\,\obe - 1505080\,\pi\,
\al^{3}\,\oal + 940320\,\al^{3}\,\opi\,\obe - 3010160\,
\pi\,\al^{3}\,\opi \\
\nonumber
 & & \mbox{} - 3295930\,\pi^{4}\,\oal - 12877972\,\al\,\pi^{3
}\,\opi - 6438986\,\pi^{3}\,\al\,\oal - 6591860\,\pi^{4}
\,\opi \\
\nonumber
 & & \mbox{} - 359200\,\al^{4}\,\opi + 7909932\,\al\,\pi^{2}
\,\obe\,\oal - 1689984\,\obe^{2}\,\al^{2}\,\beta -
20736\,\al^{2}\,\opi\,\obe^{2} \\
\nonumber
 & & \mbox{} - 69696\,\pi^{2}\,\opi\,\obe^{2} - 76032\,\pi
\,\al\,\opi\,\obe^{2} + 646560\,\obe\,\al^{3}\,\beta
 - 6234624\,\pi\,\obe^{2}\,\al\,\beta \\
\la{5.137}
 & & \mbox{} - 5749920\,\pi^{2}\,\beta\,\obe^{2}=0\,.
\end{eqnarray}
We can eliminate $\pi$ in the above equations
by defining  new variables
$\x$ and $\y$ by
\bea
\la{5.138}
&&\x:=\frac{\al}{\pi}\,,\\
\la{5.139}
&&\y:=\frac{\beta}{\opi}\,.
\eea
The side relations now assume the form (modulo non-zero factors)
\be
\la{5.140a}
S_1 := 361 + 190\,\x + 190\,\xc + 100\,\x
\,\xc - 144\,\y\,\yc=0\,,
\ee
\begin{eqnarray}
\nonumber
\lefteqn{S_2 :=  - 1663092\,\x^{2}\,\xc - 99200
\,\xc\,\x^{4} - 2527968\,\x^{2}\,\yc +
975744\,\yc^{2}\,\xc} \\
\nonumber
 & & \mbox{} - 1023264\,\yc\,\xc + 2874960\,\y
\,\yc^{2} - 4725720\,\x\,\yc - 198400\,\x
^{4} \\
\nonumber
 & & \mbox{} - 3326184\,\x^{2} - 1337120\,\x^{3} +
34848\,\yc^{2} - 3601792\,\x - 710809\,\xc \\
\nonumber
 & & \mbox{} - 2944392\,\yc + 844992\,\yc^{2}\,
\x^{2}\,\y + 1064448\,\yc^{2}\,\xc\,\x
 \\
\nonumber
 & & \mbox{} + 290304\,\yc^{2}\,\xc\,\x^{2} -
100080\,\yc\,\xc\,\x^{3} + 1728288\,\yc\,
\x^{2}\,\y \\
\nonumber
 & & \mbox{} - 667392\,\x^{2}\,\yc\,\xc +
2703384\,\yc\,\x\,\y + 1346796\,\y\,\yc -
668560\,\x^{3}\,\xc \\
\nonumber
 & & \mbox{} - 450720\,\x^{3}\,\yc - 1800896\,\x\,\xc
+ 10368\,\x^{2}\,\yc^{2} + 38016\,\x\,\yc^{2} \\
\nonumber
 & & \mbox{} - 1445316\,\x\,\yc\,\xc + 357120\,
\yc\,\x^{3}\,\y  \\
\la{5.141a}
& & \mbox{}+3117312\,\yc^{2}\,\x\,\y - 1421618=0\,,
\end{eqnarray}
\begin{eqnarray}
\nonumber
\lefteqn{S_3:= 2343618\,\x^{2}\,\xc + 89800\,
\xc\,\x^{4} - 2636184\,\x^{2}\,\yc +
975744\,\yc^{2}\,\xc} \\
\nonumber
 & & \mbox{} - 2575089\,\yc\,\xc + 2874960\,\y
\,\yc^{2} + 3295930 - 4926492\,\x\,\yc + 179600
\,\x^{4} \\
\nonumber
 & & \mbox{} + 4687236\,\x^{2} + 1505080\,\x^{3} +
34848\,\yc^{2} + 6438986\,\x + 1647965\,\xc \\
\nonumber
 & & \mbox{} - 3068538\,\yc + 844992\,\yc^{2}\,\x^{2}\,
\y + 1064448\,\yc^{2}\,\xc\,\x
 \\
\nonumber
 & & \mbox{} + 290304\,\yc^{2}\,\xc\,\x^{2} -
343080\,\yc\,\xc\,\x^{3} - 2094912\,\yc\,
\x^{2}\,\y \\
\nonumber
 & & \mbox{} - 2020092\,\x^{2}\,\yc\,\xc -
4456692\,\yc\,\x\,\y - 3122460\,\y\,\yc +
 752540\,\x^{3}\,\xc \\
\nonumber
 & & \mbox{} - 470160\,\x^{3}\,\yc + 3219493\,\x\,\xc
+ 10368\,\x^{2}\,\yc^{2} + 38016\,\x\,\yc^{2} \\
\la{5.142a}
 & & \mbox{} - 3954966\,\x\,\yc\,\xc - 323280\,
\yc\,\x^{3}\,\y + 3117312\,\yc^{2}\,\x\,\y=0\,.
\end{eqnarray}
In order to study the possible solutions of the polynomial systems
that appear from the side relations, we
 apply the  procedure {\tt gsolve} which is part of the Maple package {\tt grobner}
 \cite{czat}.
 This procedure
 computes a collection of reduced
lexicographic Gr\"obner bases corresponding to  a
set of polynomials. The system corresponding to the set is
first subdivided by factorization. Then a variant of Buchberger's
algorithm which factors all intermediate results is applied to each
subsystem.
The result is a list of reduced subsystems whose roots
are those of the original system, but whose variables have been
successively eliminated and
  separated as far as possible. This means that instead of trying to find a Gr\"obner
basis, the package attempts to factor the polynomials that
 form the system after  each step of the reduction algorithm.
 In the algorithm, the variables $\x$, $\y$ and
their complex conjugates, $\xc$ and $\yc$, are treated as
independent variables. In the subsequent analysis we use the
fact that they are complex conjugates of each other.

Applying  {\tt gsolve} to the set of equations formed by
$S_2=0$, $S_3=0$ (cf. (\ref{5.141a}) and (\ref{5.142a})),
their complex conjugates, and $S_1=0$ (cf. (\ref{5.140a})), we find the the only
possible solution for which $\x \neq 0$ and $\y \neq 0$ is given by
\be
\la{5.140}
324\y \yc-1=0\,,\qquad 6\x+11=0\,,\qquad 6\xc+11=0\,.
\ee
By substituting (\ref{5.140}) into any of the previous
expressions for $\Phi_{11}$  we find  that $\Phi_{11}=0$. Using this and
 $\pi=-\al \,6/11$ in $V_{11}$ one gets
\be
\la{salvation}
-1761\obe\al\oal+5\al^2\oal+3267\beta\obe^2-5445\beta\al\obe=0\,.
\ee
It is easy to verify that (\ref{salvation}) and the first equation in
 (\ref{5.140}), now given in the form $1089\beta\obe-\al\oal=0$,
imply that $\al=\beta=0$.

Let us consider first the cases in which each one
 of the denominators $d_1$,  $d_2$,  $d_3$,
 $d_4$ and  $d_5$, given respectively by (\ref{5.128}), (\ref{5.130}),
(\ref{5.133}), (\ref{5.134}) and (\ref{den4.5}), is zero.\\

$\bf{(i)\;d_1=0}$

From (\ref{5.128}) we have, in terms of the variables $\x$ and $\y$:
\be
\la{den4.1}
820\,\x^{2} + 1508\,\x
+ 5448\,\yc\,\x - 95 + 10236\,\yc=0\,.
\ee
Since the numerator of (\ref{5.127}) must also vanish, we have
\bea
\non
&& 2680\,\x^{3} + 1340\,
\x^{3}\,\xc - 1908\,\yc\,\xc\,\x^{2
} + 7876\,\x^{2}\,\xc \\
\non
 & & \mbox{} - 360\,\yc\,\x^{2} + 15752\,\x^{2}
 - 4824\,\yc\,\y\,\x^{2} + 30814\,\x -
1344\,\yc\,\x \\
\non
 & & \mbox{} + 15407\,\x\,\xc - 19188\,\yc\,
\y\,\x - 7296\,\yc\,\xc\,\x + 20064
 + 10032\,\xc \\
\la{num4.1}
 & & \mbox{} - 19008\,\yc\,\y - 6963\,\yc\,
\xc - 1254\,\yc=0\,.
\eea
Applying
{\tt gsolve} to the set of equations consisting
of (\ref{den4.1}), (\ref{num4.1}), their complex conjugates,
and $S_1=0$ (cf. eq. (\ref{5.140a})) we find that all possible solutions
require that $\y=0$.\\

$\bf{(ii)\;d_2=0}$

In this case, from (\ref{5.130})), we have
\be
\la{den4.2}
230\,\x + 437 - 372\,\yc=0\,.
\ee
This implies that the numerator
of (\ref{5.129}) must  vanish, so we have
\bea
\non
&&350\,\x^{2}\,\xc
+ 700\,\x^{2} + 1325\,\x\,\xc - 36\,\yc\,\x + 2650\,\x+ 2508  \\
\non
 & & \mbox{} - 450\,\yc\,\xc\,\x - 1260\,\yc\,
\y\,\x - 825\,\yc\,\xc - 66\,\yc - 2376\,\yc\,\y  \\
\la{num4.2}
 & & \mbox{} + 1254\,\xc=0\,.
\eea
Applying {\tt gsolve}
to the set of equations consisting of
(\ref{den4.2}), (\ref{num4.2}), their complex conjugates, and $S_1=0$
we find again  that all solutions require that $\y=0$.\\

$\bf{(iii)\;d_3=0}$

Eq.  (\ref{5.133}) now gives
 \be
  \la{5.142}
\Phi_{11}=-2\pi\opi(\x+2)(3\y+\xc)\,. \ee By subtracting
(\ref{5.142}) from its complex conjugate we obtain \be \la{5.143}
E_1:=-6\y-3\x\y-2\xc+6\yc+3\yc\xc+2\x=0\,. \ee Subtracting
(\ref{5.142}) from (\ref{5.127}) and taking the numerator, we get
\bea \nonumber
&&E_2:=246\y\x^2+268\x^2+216\x^2\xc^2+477\x\y+1152\yc\x\y\\
\nonumber
&&-36\yc\x+1066\x+354\yc\x\xc+2232\y\yc+518\xc-66\yc\\
\la{5.144}
&&+711\yc\xc-30\y+692\x\xc+1056=0\,.
\eea
An additional side relation is obtained by subtracting the complex conjugate
of (\ref{5.142}) from (\ref{5.129}) and taking the numerator of the resulting
expression:
\bea
\nonumber
&&E_3:=144\y\yc+2691\x\y+2622\y+144\yc\x\y+690\y\x^2\\
\nonumber
&&-120\x^2\xc-428\x\xc+81\y\x-380\xc-2508+66\yc-2650\x\\
\la{5.145}
&&+36\yc\x-700\x^2+78\y\x\xc=0\,.
\eea
Applying {\tt grobner} to the set consisting of $E_2$, its complex conjugate,
$E_1$, $E_3$ and $S_1$ we find that this system admits no solution. \\

$\bf{(iv)\;d_4=0}$

When the denominator of (\ref{5.132}), given by (\ref{5.134}), is zero, its
numerator must be zero, implying that $\al=\obe(443/3)$. This, on the
other hand implies  immediately, from $S_1=0$ (cf. (\ref{5.108}),
that $\beta=0$.\\

$\bf{(v)\;d_5=0}$

When the denominator of (\ref{5.135}) is zero, its numerator must be zero too. Thus,
we get
\be
\la{5.145b}
(11+6\x)^2(38+19\xc`12\yc\xc+10\x\xc+20\x-36\y\yc)=0\,,
\ee
\be
\la{5.145c}
308\x+513+20\x^2-528\yc-288\yc\x=0\,.
\ee
Applying {\tt gsolve} to
the polynomial system defined by the system of polynomials defined
by (\ref{5.145b}), (\ref{5.145c}), their complex conjugates, and
$S_1=0$, we find that there are no possible solutions.

The case $\al=\beta=\pi=0$, which leads to Theorem \ref{teo-4.2},
 was considered in \cite{and96} and
results in  a contradiction to the hypothesis $\phi_2 \neq 0$ .
 The more general case,
$\al\pi\beta=0$, also leads to a contradiction. The proof, found in
\cite{sas97}, is tedious but straightforward, and will not be presented here.
 Thus, the necessary conditions $I-VI$
for the validity of Huygens' principle imply that
we must have $H_{ab}:=A_{[a,b]}=0$.

The last step of the proof requires the use of the following lemma
\cite{gun52,mcl82}:
\begin{lemma}
Every scalar wave equation of the form (\ref{1.1}) for which
$A_{[i,j]}=0$, is related by a trivial transformation to one for which $A_i=0$.
\end{lemma}

From necessary condition $I$ (cf. eq. (\ref{conI})) we now have
$B=R/6$, so the wave equation is conformally invariant. Thus, the
Main Theorem is proved. \vspace{1cm}
  \pagebreak
\begin{center}
{\Large Appendix}
\end{center}

In this Appendix we give the Newman-Penrose field equations  and
commutation relations referred along the paper. We note that many
lists in the literature contain typographic mistakes, or use
different conventions.

\begin{center}
 {\large \bf Bianchi identities}
\end{center}

\[
\left.\begin{array}{l}
{(NP1)}\;\; \D\rho - \ode\kappa=\rho^{2} + \sigma
\osi + (\epsilon + \oeps)\rho - \oka\tau - (3\al
 + \obe - \pi)\kappa + \Phi_{00},\\
{(NP2)}\;\; \D\sigma - \delta\kappa=
  (\rho + \orho)\sigma + (3\epsilon - \oeps)\sigma - (
\tau - \opi + \oal + 3\beta)\kappa + \Psi_0
,
\\
{(NP3)}\;\; \D\tau - \Delta\kappa=
  (\tau + \opi)\rho + (\otau + \pi)\sigma + (\epsilon -
\oeps)\tau - (3\gamma + \oga)\kappa + \Psi_1
 + \Phi_{01}
,\\
{(NP4)}\;\; \D\al - \ode\epsilon=
  (\rho + \oeps - 2\epsilon)\al + \beta\osi - \obe\epsilon -
\kappa\lambda - \oka\gamma + (\epsilon + \rho)\pi + \Phi_{10}
,\\
{(NP5)}\;\; \D\beta - \delta\epsilon=
  (\al + \pi)\sigma + (\orho - \oeps)\beta - (\mu + \gamma)\kappa +
(\opi - \oal)\epsilon + \Psi_1
,
\\
(NP6)\;\; \D\gamma - \Delta\epsilon=(\tau
 + \opi)\al + (\otau + \pi)\beta - (\epsilon +
\oeps)\gamma - (\gamma + \oga)\epsilon
 + \tau\pi - \nu\kappa \\ \qquad \qquad \;+ \Psi_2 - \Lambda + \Phi_{11},
\\
{(NP7)}\;\; \D\lambda - \ode\pi=
  \rho\lambda + \osi\mu + \pi^{2} + (\al - \obe)\pi
 - \nu\oka + (\oeps - 3\epsilon)\lambda + \Phi_{20}
,\\

(NP8)\;\; \D\mu - \delta\pi=
 \orho\mu + \sigma\lambda + \pi\opi - (\epsilon +
 \oeps)\mu - (\oal - \beta)\pi - \nu\kappa + \Psi_2 + 2\Lambda
,\\
(NP9)\;\; \D\nu - \Delta\pi=
  (\otau + \pi)\mu + (\tau + \opi)\lambda + ({\gamma}
 - \oga)\pi - (3\epsilon + \oeps)\nu + \Psi_3
 + \Phi_{21}
,\\
(NP10)\;\;
  \Delta\lambda - \ode\nu=(\oga - 3\gamma
 - \mu - \omu)\lambda + (3\al + \obe + \pi - \otau
)\nu - \Psi_4
,\\
(NP11)\;\; \delta\rho - \ode\sigma=
  (\oal + \beta)\rho - (3\al - \obe)\sigma + (
\rho - \orho)\tau + (\mu - \omu)\kappa - \Psi_1
 + \Phi_{01},\\
(NP12)\;\; \delta\al - \ode\beta=\mu\rho - \sigma\lambda +
\al\oal + \beta\obe - 2\al\beta + (
\rho - \orho)\gamma
  + (\mu - \omu)\epsilon \\ \qquad \qquad \;
 - \Psi_2 + \Lambda + \Phi_{11}\,,\\
(NP13)\;\; \delta\lambda - \ode\mu=
 (\rho - \orho)\nu + (\mu - \omu)\pi + (\al +
 \obe)\mu + (\oal - 3\beta)\lambda - \Psi_3
 + \Phi_{21}
,\\
(NP14)\;\; \delta\nu - \Delta\mu=
 \mu^{2} + \lambda\ola + (\gamma + \oga)\mu -
\onu\pi + (\tau - \oal - 3\beta)\nu + \Phi_{22}
,\\
(NP15)\;\; \delta\gamma - \Delta\beta=
  (\tau - \oal - \beta)\gamma + \mu\tau - \sigma\nu - \epsilon
\onu - (\gamma - \oga - \mu)\beta + \al\ola +
\Phi_{12}
,\\
(NP16)\;\; \delta\tau - \Delta\sigma=
\mu\sigma + \ola\rho + (\tau - \oal + \beta)\tau
 - (3\gamma - \oga)\sigma - \kappa\onu + \Phi_{02}
,\\
(NP17)\;\; \Delta\rho - \ode\tau=
   - \rho\omu - \sigma\lambda + (\gamma + \oga)\rho -
(\otau + \al - \obe)\tau + \nu\kappa - \Psi_2 - 2\Lambda
,
\\
(NP18)\;\; \Delta\al - \ode\gamma=
  (\epsilon + \rho)\nu - (\tau + \beta)\lambda + (\oga
 - \omu)\al + (\obe - \otau)\gamma - \Psi_3\,.
\end{array}\right.
$$
 \vglue 1cm
\begin{center}{\large \bf Ricci identities}\end{center}
$$
\left.\begin{array}{l}
(NP19)\;\; \ode\Psi_0 - \D\Psi_1 +
\D\Phi_{01} - \delta\Phi_{00}=(4\al
 - \pi)\Psi_0
 - 2(2\rho + \epsilon)\Psi_1 \\ \qquad \qquad \;+ 3\kappa\Psi_2
 + (\opi - 2\oal - 2\beta)\Phi_{00}
  \mbox{} + 2(\epsilon + \orho)\Phi_{01} + 2\sigma\Phi_{10}
 - 2\kappa\Phi_{11} - \oka\Phi_{02},
\end{array}\right.
$$
$$
\left.\begin{array}{l}
(NP20)\;\; \Delta\Psi_0 -
\delta\Psi_1 + \D\Phi_{02} - \delta\Phi_{01}=(4\gamma
 - \mu)\Psi_0
 - 2(2\tau + \beta)\Psi_1 + \\ \qquad \qquad \; 3\sigma\Psi_2
 - \ola\Phi_{00} + 2(\opi - \beta)\Phi_{01} + 2
\sigma\Phi_{11}
 + (2\epsilon - 2\oeps + \orho)\Phi_{02}
 - 2\kappa\Phi_{12}
\,,
\end{array}\right.
$$
$$
\left.\begin{array}{l}
(NP21)\;\; 3\ode\Psi_1 -
 3\D\Psi_2 + 2\D\Phi_{11} - 2\delta\Phi_{10} + \ode\Phi_{01}
  - \Delta\Phi_{00}=3\lambda\Psi_0 - 9\rho
\Psi_2  \\ \qquad \qquad \;
+ 6(\al - \pi)\Psi_1 + 6\kappa\Psi_3
  + (\omu - 2\mu - 2\gamma - 2\oga)
\Phi_{00} + (2\al + 2\pi + 2\otau)\Phi_{01} \\ \qquad \qquad \;
  + 2(\tau - 2\oal + \opi)\Phi_{10}
 + 2(2\orho - \rho)\Phi_{11} + 2\sigma\Phi_{20} -
\osi\Phi_{02}
 - 2\oka\Phi_{12} - 2\kappa\Phi_{21}
\,,
\end{array}\right.
$$
$$
\left.\begin{array}{l}
(NP22)\;\; 3\Delta\Psi_1 -
 3\delta\Psi_2 + 2\D\Phi_{12} - 2\delta\Phi_{11}+ \ode\Phi_{02}
  - \Delta\Phi_{01}=3\nu\Psi_0 + 6(\gamma -
 \mu)\Psi_1 \\ \qquad \qquad \;
- 9\tau\Psi_2 + 6\sigma\Psi_3 -
\onu\Phi_{00}
  + 2(\omu - \mu - \gamma)\Phi_{01} - 2
\ola\Phi_{10} + 2(\tau + 2\opi)\Phi_{11} \\ \qquad \qquad \;
 + (2\al + 2\pi + \otau - 2\obe)
\Phi_{02} + (2\orho - 2\rho - 4\oeps)\Phi_{12}
 + 2\sigma\Phi_{21}  - 2\kappa\Phi_{22}\,,
\end{array}\right.
$$
$$
\left.\begin{array}{l}
(NP23)\;\; 3\ode\Psi_2 -
3\D\Psi_3 + \D\Phi_{21} - \delta\Phi_{20} + 2
\ode\Phi_{11}
- 2\Delta\Phi_{10}=6\lambda\Psi_1 -
9\pi\Psi_2 \\ \qquad \qquad \;
+ 6(\epsilon - \rho)\Psi_3
+ 3\kappa\Psi_4 - 2
\nu\Phi_{00}
 + 2(\omu - \mu - 2\oga)\Phi_{10}
 + (2\pi + 4\otau)\Phi_{11} \\ \qquad \qquad \;
 + (2\beta + 2\tau + \opi - 2\oal)
\Phi_{20} - 2\osi\Phi_{12} + 2(\orho - \rho - \epsilon)\Phi_{21}
- \oka\Phi_{22} + 2\lambda\Phi_{01}
,\end{array}\right.
$$
$$ \left.\begin{array}{l}
(NP24)\;\; 3\Delta\Psi_2 -
3\delta\Psi_3 + \D\Phi_{22} - \delta\Phi_{21} + 2
\ode\Phi_{12}
 - 2\Delta\Phi_{11}=6\nu\Psi_1
- 9\mu\Psi_2 \\ \qquad \qquad \;
+ 6(\beta - \tau)\Psi_3 + 3\sigma\Psi_4
  - 2\nu\Phi_{01} - 2\onu\Phi_{10} + 2(
2\omu - \mu)\Phi_{11} + 2\lambda\Phi_{02} - \ola
\Phi_{20}\\ \qquad \qquad \;
  + 2(\pi + \otau - 2\obe)\Phi_{12}
 + 2(\beta + \tau + \opi)\Phi_{21}
  + (\orho - 2\epsilon - 2\oeps - 2\rho)
\Phi_{22}
 \,,\end{array}\right. $$
$$ \left.\begin{array}{l}
(NP25)\;\; \ode\Psi_3 - \D\Psi_4
 + \ode\Phi_{21} - \Delta\Phi_{20}=3\lambda
\Psi_2
  - 2(\al + 2\pi)\Psi_3 + (4\epsilon - \rho)\Psi_4 \\ \qquad \qquad \;
 - 2\nu\Phi_{10} + 2\lambda\Phi_{11}
 + (2\gamma - 2\oga + \omu)\Phi_{20}
 + 2(\otau - \al)\Phi_{21} - \osi\Phi_{22}
 \,,\end{array}\right. $$
$$ \left.\begin{array}{l}
(NP26)\;\; \Delta\Psi_3 - \delta\Psi_4 +
 \ode\Phi_{22}-\Delta\Phi_{21}=3\nu
\Psi_2
  - 2(\gamma + 2\mu)\Psi_3 + (4\beta - \tau)\Psi_4  \\ \qquad \qquad \;
 - 2\nu\Phi_{11} - \onu\Phi_{20} + 2\lambda
\Phi_{12}
  + 2(\gamma + \omu)\Phi_{21} + (\otau
 - 2\obe - 2\al)\Phi_{22}
 \,,\end{array}\right. $$
$$ \left.\begin{array}{l}
(NP27)\;\; \D\Phi_{11} - \delta\Phi_{10} -
 \ode\Phi_{01} + \Delta\Phi_{00} + 3
\D\Lambda=
  (2\gamma - \mu + 2\oga - \omu)\Phi_{00} \\ \qquad \qquad \;
+ (\pi -
2\al - 2\otau)\Phi_{01}
 + (\opi - 2\oal - 2\tau)\Phi_{10}
 + 2(\rho + \orho)\Phi_{11} + \osi\Phi_{02} + \sigma
\Phi_{20} \\ \qquad \qquad \;
  - \oka\Phi_{12} - \kappa\Phi_{21}
 \,,\end{array}\right. $$
$$ \left.\begin{array}{l}
(NP28)\;\; \D\Phi_{12} - \delta\Phi_{11}
 - \ode\Phi_{02}
+ \Delta\Phi_{01} + 3
\delta\Lambda=
  (2\gamma - \mu - 2\omu)\Phi_{01} +\\ \qquad \qquad \;
 \onu
\Phi_{00} - \ola\Phi_{10} + 2(\opi - \tau)\Phi_{11}
  + (\pi + 2\obe - 2\al - \otau)\Phi_{02} \\ \qquad \qquad \;
+
(2\rho + \orho - 2\oeps)\Phi_{12} + \sigma
\Phi_{21} - \kappa\Phi_{22}
\,,\end{array}\right. $$
$$ \left.\begin{array}{l}
(NP29)\;\; \D\Phi_{22} - \delta\Phi_{21}
 - \ode\Phi_{12} + \Delta\Phi_{11} + 3
\Delta\Lambda=\nu\Phi_{01}
  + \onu\Phi_{10} - 2(\mu + \omu)\\ \qquad \qquad \;
\Phi_{11} - \lambda\Phi_{02} - \ola\Phi_{20}
  + (2\pi - \otau + 2\obe)\Phi_{12}
 + (2\beta - \tau + 2\opi)\Phi_{21} \\ \qquad \qquad \;
 + (\rho + \orho - 2\epsilon - 2\oeps)\Phi_{22}
\,.\end{array}\right. $$
\begin{center}
\vglue 1cm
{\large \bf NP commutation relations}
\end{center}
\[
\left.\begin{array}{l}
\ode\delta-\delta\ode=(-\mu+\omu)\D+(-\rho+\orho)\Delta+
(\al-\obe)\delta+(-\oal+\beta)\ode\,,\\
\ode\Delta-\Delta\ode=-\nu\D+(\otau-\al-\obe)
\Delta+\lambda\delta+(\omu+\gamma-\oga)\ode\,,\\
\ode\D-\D\ode=(\al+\obe-\pi)\D+\oka\Delta-\osi\delta-
(\rho-\epsilon+\oeps)\ode\,,\\
\Delta\D-\D\Delta=(\gamma+\oga)\D+(\epsilon+\oeps)\Delta
-(\otau+\pi)\delta-(\tau+\opi)\ode\,.
\end{array}\right.
\]

 \vglue 1 cm
\noindent
 \Large
\end{document}